\documentstyle[aps,epsf]{revtex}  
\begin{document} 
\def\Astrobiologia{Astrobiolog\'\i{}a}
\def\Astrofisica{Astrof\'\i{}sica}
\def\Paris{Par\'\i{}s}
\def\Perez{P\'erez}
\def\Fisica{F\'\i{}sica}
\def\Torrejon{Torrej\'on}
\title{ 
Effective potential for the reaction-diffusion-decay system
} 
\author{ 
David Hochberg$^{+,\dag}$, Carmen Molina--\Paris,$^{++,}$
\thanks{Corresponding author:
Theoretical Division T-8, MS B285, Los Alamos National Laboratory,
Los Alamos, NM 87545;
Phone: (505)665-8824;
FAX: (505)665-3003;
Electronic Address: carmen@t6-serv.lanl.gov},
Juan \Perez--Mercader$^{+++,\dag}$, and Matt Visser$^{++++}$ 
} 
\address{ 
$^{+,+++}$Laboratorio de \Astrofisica\ Espacial y \Fisica\ 
Fundamental, Apartado 50727, 28080 Madrid, Spain\\ 
$^{\dag}$Centro de \Astrobiologia,
CSIC/INTA, Ctra. Ajalvir, Km. 4, 28850 \Torrejon, Madrid, Spain\\
$^{++}$Theoretical Division, Los Alamos National Laboratory, 
Los Alamos, New Mexico 87545, USA\\ 
$^{++++}$Physics Department, Washington University, 
Saint Louis, Missouri 63130-4899, USA} 
\date{17 August 1999; \LaTeX-ed \today} 
\maketitle 

\bigskip

{\small 

{\bf Abstract:} In previous work we have developed a general method
for casting stochastic partial differential equations (SPDEs) into a
functional integral formalism, and have derived the one-loop effective
potential for these systems.  In this paper we apply the same
formalism to a specific field theory of considerable interest, the
reaction-diffusion-decay system.  When this field theory is subject to
white noise we can calculate the one-loop effective potential (for
arbitrary polynomial reaction kinetics) and show that it is one-loop
ultraviolet renormalizable in 1, 2, and 3 space dimensions.  For
specific choices of interaction terms the one-loop renormalizability
can be extended to higher dimensions. We also show how to include the
effects of fluctuations in the study of pattern formation away from
equilibrium, and conclude that noise affects the stability of the
system in a way which is calculable.
 
\bigskip


\bigskip

} 

\newcommand{\define}{\mathop{\stackrel{\rm def}{=}}}
\newcommand{\tr}{\mathop{\mathrm{tr}}} 
\newcommand{\Tr}{\mathop{\mathrm{Tr}}}
\def\d{{\mathrm{d}}}
\def\implies{\Rightarrow}
\def\half{ {\scriptstyle{1\over2}} }
\def\quarter{ {\scriptstyle{1\over4}} }
\def\dhalf{ {\scriptstyle{d\over2}} }
\def\A{ {\cal A} }
\def\Re{ {\mathbf Re} }
\def\Im{ {\mathbf Im} }
\def\SIZE{1.00}
\section{Introduction}

Geometrical patterns are ubiquitous: From galaxies to living systems,
examples abound where a particular spatial distribution of some
material is preferred versus others out of a seemingly unlimited
variety. In many cases, these patterns are successfully described by
systems of coupled parabolic non-linear partial differential
equations. This is the case, for example, in chemical kinetics, where
such equations summarize the space-time evolution of chemical species
diffusing and reacting in some confined geometrical region, which
makes its presence felt in the boundary conditions for the problem. In
this way chemical kinetics helps one to understand leopard spots,
zebra bands or the radial structure of {\em Acetabularia}.  (See
Murray~\cite{Murray}, Walgraef~\cite{Walgraef}, and Ball~\cite{ball}
for discussions of many specific examples. Useful background
references include van Kampen~\cite{van-Kampen} and
Gardiner~\cite{Gardiner}.) In these phenomena the values of ``reaction
constants'' play a role which reminds one of the role played by
coupling constants in determining the vacuum (or ground) state in a
quantum field theory undergoing spontaneous symmetry breaking, or in
the description of phase transitions in condensed matter systems. This
is not surprising, since in the presence of spatial non-equilibrium
patterns~\cite{Mori} ``it is often the case that the continuous
symmetry of the system becomes spontaneously broken''.  Because of the
above, two questions come immediately to mind: (a) what is the effect
on an existing pattern of the elimination of fast degrees of freedom?
and, (b) how do fluctuations affect the stability of an established
pattern? These two questions are, of course, formally related since
both phenomena manifest themselves through noise added to the
otherwise deterministic equations describing the formation of these
patterns. Providing answers to these questions opens the door to the
study of complex phenomena where either there is no precise explicit
knowledge of many of the microscopic details, or else where unexpected
external perturbations and disturbances show up, and one is
nevertheless interested in an explicit understanding of the system at
long wavelengths. An example is in population ecology, where the
presence of illegal hunting, or the accidental introduction of some
{(apparently minor)} contaminant, can produce major ecological shifts.

The most appropriate tool to study patterns of symmetry is, of course,
the notion of a potential. If some notion of potential is available,
then analysis of its extrema leads to the identification of the stable
and metastable vacua for the system. From there, given the values of
the couplings in the system, one determines the ground state for the
system. This analysis is performed in equilibrium, but the problem in
pattern formation, or in reaction-diffusion systems, is that these
systems are away from equilibrium, often far away from equilibrium,
and the usual notion of ``potential'', ``vacuum state'', and allied
concepts are no longer available. Recently, however, we have
introduced~\cite{HMPV-spde} a notion of ``effective potential'' which
generalizes the standard (Quantum Field Theory) notion of an effective
potential to systems away from equilibrium. The generalization (as
discussed below) is such that a very clean and clear parallel can be
established with the situation in Quantum Field Theory, and a
potential is constructed which has two major and distinct pieces: a
``classical'' contribution and a ``fluctuation'' contribution. The
``classical'' contribution plus the ``fluctuation'' contribution
determine the ``vacuum state'' of the system, and therefore this
effective potential allows the calculation of the effects of the
fluctuations on the ground state of the system. { The ``minimalist
formalism'' we developed in references~\cite{HMPV-spde} is to be
contrasted with the traditional Martin-Siggia-Rose (MSR) formalism
with its extra unphysical conjugate
fields~\cite{MSR,De-Dominicis-Peliti}. The minimalist formalism is an
extension and outgrowth of the Onsager-Machlup
approach~\cite{Onsager-Machlup,Eyink1}, and exhibits similarities to
the analysis of Crisanti and Marconi~\cite{Crisanti-Marconi}.}

The fluctuation dependent piece of the potential involves integrations
over the frequency and momentum domains. These integrals require the
introduction of a cutoff, which, through a Wilsonian-style procedure,
leads to a scale-dependence of the parameters of the
reaction-diffusion-decay system and therefore has an effect on the
type of instability (and associated patterns) which controls the
behavior of the system. The full ``effective potential'' then
constitutes a superb tool to incorporate the effects of
non-linearities and fluctuations in the patterns produced by systems
away from equilibrium of the reaction-diffusion type.

In this paper, after a brief discussion of the ``effective potential''
in general, and a description of how the noise amplitude plays the
role of a loop counting parameter, we specialize to the case of
reaction-diffusion-decay systems. The ``potential'' is calculated for
$d=0,1,2,$ and $3$ spatial dimensions, and the special case of
monomial interactions is given separate treatment. We also provide
specialized discussions for higher dimensions and general colored
noises. Then, as an application of the previous results, we study the
effects of Gaussian white noise in both Hopf and Turing bifurcations
by computing appropriate quantities without fluctuations and with
fluctuations. We find that the effect of noise is to shift the
symmetric states of the system, {\it as well as} to change the nature
of the linear instabilities that may occur as perturbations around
these new states. We end by offering some conclusions.

{ To fix the notation:} In a companion paper~\cite{HMPV-spde}, we
discussed classical field theories subject to {additive} stochastic
noise $\eta(\vec x,t)$ described by the equation
\begin{equation}
D \phi(\vec x,t) = F[\phi(\vec x,t)] + \eta(\vec x,t)
\; .
\label{E:spde}
\end{equation}
Here $D$ is any linear differential operator, involving arbitrary time
and space derivatives, that does {\em not} explicitly involve the
field $\phi$.  The function $F[\phi]$ is any forcing term, generally
nonlinear in the field $\phi$. These stochastic partial non-linear
differential equations (SPDEs) can be studied using a functional
integral formalism which makes manifest the deep connections with
quantum field theories (QFTs). We showed that if the noise is
translation-invariant and Gaussian, it is possible to split its
two-point function into an {\em amplitude} ${\A}$ and a {\em shape}
function $g_2(x,y)$, as follows
\begin{equation}
G_\eta(x,y) \define \A \; g_2(x-y)
\; ,
\end{equation}
with the {\em convention} that
\begin{equation}
\label{E:convention}
\int \d^d \vec x\; \d t \;  g_2^{-1}(\vec x,t)
\; = \; 1
\; = \; \tilde g_2^{-1}(\vec k=\vec 0,\omega=0)
\; .
\end{equation}
Then the one-loop effective potential for the SPDE is~\cite{HMPV-spde}
\begin{eqnarray}
\label{E:general}
{\cal V}[\phi;\phi_0] &=& 
\half F^2[\phi]
+ \half \A \int {\d^d \vec k \; \d \omega\over (2\pi)^{d+1}}
\ln  
\left[ 1 + {\tilde g_2{}(\vec k,\omega)
F[\phi] {\delta^2 F\over\delta\phi\;\delta\phi} 
\over
\left( D^\dagger(\vec k,\omega)  - {\delta F\over \delta\phi}^\dagger \right)
\left( D(\vec k,\omega) - {\delta F\over \delta\phi} \right)}
\right]
\nonumber\\
&&- \left( \phi \to \phi_0 \right)
+ O(\A^2)\; .
\label{E:p1spde}
\end{eqnarray}
Here $\phi_0$ is any convenient background field.  The above result is
qualitatively similar to the one-loop effective potential for scalar
QFT~\cite{Weinberg,Rivers,Itzykson-Zuber,Zinn-Justin}:
\begin{eqnarray}
{\cal V}[\phi;\phi_0] &=& 
V(\phi)
+ \half \hbar \int {\d^d \vec k \; \d \omega\over (2\pi)^{d+1}}
\ln  
\left[ 1 + { {\delta^2 V\over\delta\phi\;\delta\phi} 
\over
\omega^2 + \vec k^2 + m^2
}
\right]
- \left( \phi \to \phi_0 \right)
+ O(\hbar^2) \; ,
\label{E:p1qft}
\end{eqnarray}
as can be seen by simply comparing equations (\ref{E:p1spde}) and
(\ref{E:p1qft}).  Moreover, as argued in~\cite{HMPV-spde}, this
effective potential for SPDEs inherits many of the interesting
features of the effective potential for QFTs. In particular, minima of
the effective potential for a given SPDE correspond to homogeneous and
static solutions of the stochastic equations of motion (that is,
homogeneous and static expectation values of the stochastically driven
field $\phi$, with the averaging done with respect to noise
realizations)~\cite{HMPV-spde}.

In a second paper~\cite{HMPV-kpz}, we applied this formalism to the
Kardar-Parisi-Zhang (KPZ) equation, obtaining an interesting ground
state structure (including dynamical symmetry breaking). In the
current paper we perform a similar analysis for the
reaction-diffusion-decay system described by the equation:
\begin{equation}
\label{E:version1}
\left({\partial\over\partial t} - \nu \vec\nabla^2 \right) \phi = 
P_0- \gamma \phi + P(\phi) + \eta \; .
\end{equation}
This equation can be used, for instance, as a model to describe the
dynamics and spatial distribution of the concentration of a chemical
reagent, when it is subject to both diffusion (via $\nu$) and decay
(via $\gamma$). $P(\phi)$ is some ultra-local function of the
concentration (typically a polynomial in the concentration, but not
always) and it represents the reaction
kinetics~\cite{Murray,Walgraef,ball,van-Kampen,Gardiner}.  We have included
a ``tadpole term'' $P_0$ since, as we will see, its presence is
essential for the consistency of the ultraviolet renormalization
program.  {Many examples of reaction-diffusion equations abound in the
literature. By way of example, we mention just two model equations
that the reader may wish to keep in mind. One goes under the name of
``amplitude equation'', generally complex, resulting from reducing a
(non-stochastic) reaction-diffusion equation in the vicinity of an
instability point:
\begin{equation}
\label{E:amplitude}
\partial_t A = \nu {\vec\nabla}^2 A + f(|A|^2)A 
\; ,
\end{equation}
where the diffusion coefficient can be complex: $\nu = \nu_R +
i\nu_I$.  The study of amplitude equations is very useful for
determining the basic geometry of patterns that can emerge near the
instability point~\cite{Walgraef,ball}.}  This equation clearly involves
only one field degree of freedom, but the generalization is
straightforward.  For several species of, e.g., chemical reactant the
field (concentration) $\phi$ is promoted to a vector $\phi_i(\vec
x,t)$. The diffusion coefficient and decay rate are then promoted to
matrices, the noise to a vector, and the reaction kinetic function
$P(\phi)$ to a vector-valued functional with tensorial coefficients.
\begin{equation}
\label{E:version1b}
\left( 
\delta_i{}^j \; {\partial\over\partial t} - \nu_i{}^j \; \vec\nabla^2 
\right) \phi_j = 
(P_0)_i - \gamma_i{}^j \; \phi_j + P_i(\phi_j) + \eta_i
\; .
\end{equation}
{Typical examples of this sort involving two components go under
the heading of ``activator-inhibitor'' models of e.g., biological
pattern formation, a noise-free example of which is provided by the
Gierer-Meinhardt mechanism:
\begin{eqnarray}
\label{E:GM}
\partial_t A &=& D_A {\vec\nabla}^2 A + k_1-k_2A + \frac{k_3A^2}{B}
\; ,
\\
\partial_t B &=& D_B {\vec \nabla}^2 A + k_4A^2-k_5B
\; .
\end{eqnarray}
Here $\phi_1 = A$ is the activator and $\phi_2 = B$ 
is the inhibitor. The $k_i$ are reaction constants.} 

Finally, we note that these systems provide a viable framework for the
notion of ``self-organizing systems''.

{The remainder of this paper is organized as follows. In Section II,
following the general procedure developed in~\cite{HMPV-spde}, we
construct the complete one-loop effective potential for the class of
reaction-diffusion equations in (\ref{E:version1}). Once this is done,
we carry out a detailed analysis of the resulting potential as a
function of spatial dimension for $d=0,1,2,3$. Special attention is
also paid to monomial interactions, the case of higher spatial
dimensions, and the case of correlated Gaussian noise. In Section III
we turn to a discussion of the impact that noise and fluctuations can
have on the onset of instabilities and pattern formation. We conclude
in Section IV with a discussion of our results. Certain technical
issues having to do with functional Jacobian determinants are
collected in Appendixes A and B. The general Feynman rules for the
reaction-diffusion equation (\ref{E:version1}) are presented in
Appendix C. An integral needed in the computation of the one-loop
effective potential is calculated in Appendix D. Finally, it should be
borne in mind that the main body of the paper adopts the so-called
{\em Stratonovich calculus}---Appendix E indicates the changes that are
required if one wishes to consider the so-called {\em Ito calculus}.}

\section{Effective Potential: Reaction-diffusion-decay systems}

{ We begin this section by computing the one-loop effective
potential associated to the class of reaction-diffusion equations
(\ref{E:version1}).  We first carry out the computations for arbitrary
Gaussian noise and arbitrary spatial dimension, only later
specializing to white Gaussian noise and examining the particular
features of the effective potential for dimensions $d=0,1,2,3$.}

To avoid unnecessary clutter, let us re-write equation (\ref{E:version1}) as
\begin{equation}
\label{E:version2}
\left({\partial\over\partial t} - \nu \vec\nabla^2\right) \phi = 
P(\phi) + \eta\; .
\end{equation}
Here $P(\phi)$ is an arbitrary polynomial in the field $\phi$. Any
tadpole contribution $P_0$, as well as any decay term $-\gamma \phi$,
have now for convenience been subsumed into $P(\phi)$.

To apply the general analysis provided in~\cite{HMPV-spde}, which led
to equations (\ref{E:spde})-(\ref{E:p1spde}) above, 
to a homogeneous and static background
field, $\phi(\vec x,t) = \mathrm{constant}$,
we simply make the identifications
\begin{equation}
F[\phi] \to P(\phi); 
\qquad
{\delta F\over\delta\phi(x)} \to P'(\phi);
\qquad
{\delta^2 F \over \delta\phi(x) \delta\phi(y)}
\to P''(\phi)
\; ,
\end{equation}
as follows by comparing equations (\ref{E:spde}) and (\ref{E:version2}).
At tree-level (zero-loop) in the loop-counting parameter $\A$, the
equations of motion become~\cite{HMPV-spde}
\begin{equation}
{\delta F\over\delta\phi}^\dagger F[\phi] = J
\qquad \to \qquad 
P'(\phi) P(\phi) = J
\; .
\end{equation}
This is a polynomial equation for $\phi$ and therefore has a finite
number of roots.  In particular, for $J=0$ let $\phi_0$ be {\em one}
of the roots of $P'(\phi)P(\phi)=0$. This polynomial always has at
least one real root. [Proof: let $P(\phi)$ be of degree $n$, then
$P^2(\phi)$ is of degree $2n$, and $P'(\phi)P(\phi)$ is of degree
$2n-1$, which is always odd.  Thus $P'(\phi)P(\phi)$ must cross the
abscissa at least once, so there must be at least one real zero.]

The zero-loop effective potential is now
\begin{equation}
{\cal V}_{\mathrm zero-loop}[\phi;\phi_0] = 
\half \left[P^2(\phi) -  P^2(\phi_0) \right] \; .
\end{equation}
This zero-loop effective potential is both a generalization (because
it includes the effects of non-linearities) and a specialization
(because it treats static fields) of the Onsager-Machlup action for
stochastic mechanics~\cite{HMPV-spde,Onsager-Machlup,Eyink1}.

We start the one-loop computation by noting that for the
reaction-diffusion-decay system the linear differential operator
needed is given by
\begin{equation}
D - {\delta F\over\delta \phi} = 
\partial_t - \nu \vec \nabla^2 - P'(\phi)
\qquad \to \qquad -i\omega + \nu \vec k^2 - P'(\phi) \; ,
\end{equation}
in configuration and Fourier variables. We then have for the adjoint
 quantity
\begin{equation}
D^\dagger - {\delta F\over\delta \phi}^\dagger = 
-\partial_t - \nu \vec \nabla^2 - P'(\phi)
\qquad \to \qquad +i\omega +\nu \vec k^2 - P'(\phi) \; ,
\end{equation}
so that
\begin{equation}
\left(D^\dagger -{\delta F\over\delta\phi}^\dagger\right)
\left(D - {\delta F\over\delta\phi}\right)
 = -\partial_t^2 + [\nu \vec \nabla^2 + P'(\phi)]^2
\qquad \to \qquad +\omega^2 +[\nu \vec k^2 - P'(\phi)]^2 \; .
\end{equation}
Using the previous equations, which are specializations of the general
ones presented in~\cite{HMPV-spde}, the one-loop effective potential is
given by
\begin{eqnarray}
\label{E:rdd-general-noise}
{\cal V}[\phi;\phi_0] &=& 
\half P^2(\phi)
+ \half \A \int {\d^d \vec k \; \d \omega\over (2\pi)^{d+1}}
\ln  
\left[ 1 + {\tilde g_2{}(\vec k,\omega) 
 P(\phi) P''(\phi)
\over 
\omega^2 + [\nu \vec k^2 - P'(\phi)]^2}
\right]
- \left( \phi \to \phi_0 \right)
+ O(\A^2) \; .
\end{eqnarray}
Equivalently
\begin{eqnarray}
{\cal V}[\phi;\phi_0] &=& 
\half P^2(\phi)
+ \half \A \int {\d^d \vec k \; \d \omega\over (2\pi)^{d+1}}
\ln  
\left[ {
\omega^2 + [\nu \vec k^2 - P'(\phi) ]^2 + 
\tilde g_2{}(\vec k,\omega) 
P(\phi) P''(\phi) 
\over 
\omega^2 + [\nu \vec k^2 - P'(\phi) ]^2}
\right]
- \left( \phi \to \phi_0 \right)
+ O(\A^2) \; .
\label{E:p1rdd}
\end{eqnarray}
{At this stage we should be explicit about some technical details:
First, the way we have chosen to treat the functional Jacobian is
equivalent to choosing the {\em Stratonovich calculus} for the
stochastic noise~\cite{HMPV-spde}. The modifications attendant on the
choice of the {\em Ito calculus} are in some ways a simplification of
the current procedure, but in other ways lead to additional technical
complications. Appendix E discusses some features of the Ito
calculus. Second, in deriving the formulae above it has been assumed
(in performing the functional integrations) that the field $\phi$ is
unrestricted, and can take on all values from $-\infty$ to
$+\infty$. Strictly speaking, this is of course not the case if $\phi$
represents a concentration, but this is not a serious restriction. One
can deal with this either by (1) choosing the forcing term $P(\phi)$
to strongly suppress negative values of $\phi$, and then taking a
suitable limit, or more prosaicly (2) by realizing that in making the
one-loop approximation we have already assumed that fluctuations are
in some sense small, so that if we look at quadratic fluctuations
around some positive value of the background field $\phi_0$, then the
error made in letting $\delta\phi = \phi-\phi_0$ range over the entire
real line is a higher order effect [at least $O(\A^2)$.] }

This is as far as we can go {\em without making any further
assumptions about the additive noise}.  For instance, one standard
choice is {\em temporally white}, which means delta function
correlated in time, so that $\tilde g_2(\vec k,\omega)\to \tilde
g_2(\vec k)$ is a function of $\vec k$ only.  {Let us define
$X^2=[\nu \vec k^2 - P'(\phi) ]^2 + \tilde g_2{}(\vec k) P(\phi)
P''(\phi) $ and $Y^2 =[\nu \vec k^2 - P'(\phi) ]^2$. It is easy to see
from its definition that $Y^2$ is real and positive, and so is $Y$. If
$X^2$ is positive, we can make use of the
standard integral identity [$X$ and
$Y$ are positive, see~\cite{G+R}, equation (4.222.1)], namely}
\begin{equation}
\label{E:frequency}
\int_{-\infty}^{+\infty} \d \omega 
\ln \left( {\omega^2 + X^2\over \omega^2 + Y^2} \right)= 
2\pi \left( X - Y \right)\; ,
\end{equation}
to re-write equation (\ref{E:p1rdd}) as
\begin{eqnarray}
{\cal V}[\phi;\phi_0] &=& 
\half P^2(\phi) 
+ \half \A \int {\d^d \vec k  \over (2\pi)^{d}} 
\left\{
\sqrt{  [\nu \vec k^2 - P'(\phi) ]^2 
+ \tilde g_2{}(\vec k) 
P(\phi) P''(\phi) 
} 
- | \nu \vec k^2 - P'(\phi) |  
\right\}
- (\phi \to \phi_0)
+ O(\A^2) \; .
\end{eqnarray}
If $X^2$ is real and negative, we can define $X^2 = -Z^2$, with $Z$ a
real positive number.  In this case we must make use of the following
integral (see Appendix D)
\begin{equation}
\label{E:frequency2}
\int_{-\infty}^{+\infty} \d \omega 
\ln \left( {\omega^2 -Z^2 \pm i \epsilon \over \omega^2 + Y^2} \right)
= 
2\pi \left(  \pm i Z - Y \right) \; ,
\end{equation}
to get the appropriate result; here $\epsilon$ is a real, small, and
positive number.

Averaging over the $\pm$ justifies the following  prescription
\begin{equation}
\label{E:frequency3}
\int_{-\infty}^{+\infty} \d \omega 
\ln \left( {\omega^2 -Z^2 \over \omega^2 + Y^2} \right) \to 
\int_{-\infty}^{+\infty} \d \omega 
\; 
\Re\left[\ln \left( {\omega^2 -Z^2 \over \omega^2 + Y^2} \right)\right]
= 
- 2\pi Y \; .
\end{equation}
This prescription is guaranteed to preserve the reality of the
effective potential. In general we should write
\begin{eqnarray}
\label{E:integral}
{\cal V}[\phi;\phi_0] &=& 
\half P^2(\phi) 
+ \half \A \int {\d^d \vec k  \over (2\pi)^{d}} 
\left\{
\Re\left[\sqrt{  [\nu \vec k^2 - P'(\phi) ]^2 
+ \tilde g_2{}(\vec k) 
P(\phi) P''(\phi) 
} \right]
- | \nu \vec k^2 - P'(\phi) |  
\right\}
\nonumber\\
&&
- (\phi \to \phi_0)
+ O(\A^2)
\; ,
\end{eqnarray}
where we have used the $\omega$-integral to recast (\ref{E:p1rdd})
into (\ref{E:integral}).  Remembering that $\lim_{\vec k \rightarrow
\vec 0} {\tilde g_2{}}(\vec k) = 1$ [recall (\ref{E:convention})], it
is clear that there are no infrared divergences $(\vec k \rightarrow
\vec 0)$, at least for this effective potential at one-loop order. To
investigate the ultraviolet (i.e., short-distance) behaviour, it is
useful to re-express this in the form
\begin{eqnarray}
\label{E:root}
{\cal V}[\phi;\phi_0] &=& 
\half P^2(\phi) 
+ \half \A \int {\d^d \vec k  \over (2\pi)^{d}} \;
\left| \nu \vec k^2 - P'(\phi)\right| \;
\left\{
\; \Re\sqrt{ 1   
+ 
{\tilde g_2{}(\vec k) P(\phi) P''(\phi)
\over
[\nu \vec k^2 - P'(\phi) ]^2}
} 
- 1 \right\}
\nonumber\\
&&
- (\phi \to \phi_0)
+ O(\A^2)\; .
\end{eqnarray}
It is clear now that this effective potential will be finite provided
the spatial part of the noise spectrum satisfies
\begin{equation}
\label{E:finite}
\int \d^d \vec k \; 
{\tilde g_2{}(\vec k)\over |\vec k^2 +a|} < \infty \; .
\end{equation}
Thus, while the noise acts as the source of the fluctuations, it can
also serve as the regulator to keep physical quantities UV-finite, as
should be clear from (\ref{E:finite}).  Indeed, whatever
modifications one might make to the noise in the ultraviolet region
will have no consequence for the long-wavelength or hydrodynamic
limit.  The noise is intended to model fluctuations above a certain
limiting resolution length/time scale, thereby making its short
distance behavior immaterial, in so far as one is interested in
studying the long distance, long time asymptotic behavior of the
stochastic model~\footnote{ Since this point may cause some
confusion to the reader, we briefly pause to belabour it: UV
renormalizability is {\em not} the same as UV {\em finiteness} and we
do not claim that the theory has to make sense at arbitrarily short
distances. (If nothing else, in real chemical kinetics the interatomic
spacing will provide a natural UV cutoff.) What UV renormalizability
does is to sharply {\it limit} the number of relevant operators, so
that the low energy theory (long distances, large times) is guaranteed
to be relatively simple. Lack of UV renormalizability is not fatal for
a theory, but does make life considerably more complicated. This
phenomenon is known in the quantum field theory literature as
``decoupling''; see {\it e.g.,} Weinberg II~\cite{Weinberg}.}.

For definiteness, let us now take the spatial noise spectrum to be
cutoff white, {\em i.e.},
\begin{equation}
\tilde g_2(\vec k) =\tilde g_2(\vert \vec k \vert) 
=  \Theta(\Lambda - k) \; .
\end{equation}
With this choice of noise, we can Taylor expand the square root of
equation (\ref{E:root}) in the ultraviolet regime to see that there is
a divergent term proportional to $P(\phi) P''(\phi)\; \Lambda^{d-2}$,
and a subdominant divergent term proportional to $P(\phi)\; P'(\phi)
\; P''(\phi) \; \Lambda^{d-4}$. Since the classical (tree-level)
potential is just $P^2(\phi)$, to have any hope of absorbing the
infinities into the bare action we must have $d<4$.  That is: the
reaction-diffusion-decay system, subject to white noise and for {\em
any} polynomial $P(\phi)$, is one-loop ultraviolet renormalizable {\em
only} in 0,1,2, and 3 space dimensions. In $0$ space dimensions the
reaction-diffusion-decay system reduces to a Langevin reaction-decay
system which is still interesting, (see below).  In $d=1$ the
reaction-diffusion-decay system is in one-loop finite. (Strictly
speaking the claim of one-loop renormalizability also requires the
investigation of the wavefunction renormalization; this is beyond the
scope of the present paper. For related discussion
see~\cite{HMPV-sdw-rdd}.) The assertion that arbitrary polynomial
reaction kinetics can be renormalizable in low dimensions should not
(with hindsight) be alarming. After all, exactly the same thing
happens for quantum field theories in $d=2$ spacetime dimensions,
where $P(\phi)_2$ is renormalizable for arbitrary polynomials. If we
restrict the form of the polynomial occurring in the
reaction-diffusion-decay system, we can have one-loop renormalizable
theories in a higher dimensions. We will come back to this particular
point later.

To be more explicit, we expand the unrenormalized one-loop effective
potential as follows
\begin{eqnarray}
\label{E:explicit}
{\cal V}[\phi;\phi_0] &=& 
\half P^2(\phi) 
+ \half \A \int {\d^d \vec k  \over (2\pi)^{d}} \;
\sum_{n=1}^\infty \left\{ 
{\half\choose n}    
{ [P(\phi) P''(\phi)]^n
\over
[\nu \vec k^2 - P'(\phi) ]^{2n-1}} 
\right\}
- (\phi \to \phi_0)
+ O(\A^2)
\; .
\end{eqnarray}
This expansion only makes sense if $ |P(\phi) P''(\phi)| < [\nu \vec
k^2 - P'(\phi) ]^2$ for every value of {$\vert \vec k \vert$}.  This
requires both $P'(\phi) < 0$ and $|P(\phi) P''(\phi)| <[P'(\phi)]^2$.
The relevant integrals, after a rescaling, are\footnote{
See~\cite{G+R}, section (8.380.3). }
\begin{equation}
{\cal I}_{\mathrm RDD}(n,d) \define
\int_0^\infty \d x \; x^{d-1} (1+x^2)^{1-2n} =
\half B(d/2, 2n-1-d/2 ) = 
{\Gamma(d/2) \Gamma(2n-1-d/2) \over 2\Gamma(2n-1)} \; .
\end{equation}
These integrals converge
for $n>(d+2)/4$.  {For $n\leq(d+2)/4$ we should introduce
appropriate counterterms.} Of course, the one-loop effective potential
may make perfectly good sense even when this Taylor series expansion
is problematic.  {By this we simply mean that the one-loop
contribution can be finite for values of the parameters and momentum
that lie formally outside the radius of convergence of the above
series: one would then obviously not proceed by expanding and
integrating term-wise, as we have done here.}

{Finally we should remind the reader that even if the generic
reaction-diffusion-decay system is non-renormalizable for $d\geq4$,
this does not mean that such theories are completely useless. (Though
it must be admitted that the number of physically relevant examples in
four or more space dimensions is rather limited, [see for example
variants on the idea of Kaluza-Klein theory], the
non-renormalizability {\em per se} is not the issue.) All that
non-renormalizability implies is that the theory must be viewed as an
``effective field theory'' that must include many more terms in the
effective action than naively arise in the zero-loop
approximation. These new terms carry with them additional (typically
dimensionful) coupling constants, and these coupling constants govern
the range of validity of the effective field theory.
See~\cite{Weinberg} for a modern discussion of effective field
theories in the QFT context. In statistical mechanics language, the
universality class of an effective field theory is much more
complicated than would be naively deduced from the zero-loop
approximation.}

We next plunge into a discussion of the above as a function of the
{\it spatial} dimension, $d$.

\subsection{Reaction-(diffusion)-decay: $d=0$}

In $d=0$ space dimensions, {there is of course no
diffusion,} and the system reduces to a simple Langevin
reaction-decay system. We have
\begin{equation}
{\d\over\d t} \phi = 
P(\phi) + \eta \; .
\end{equation}
{Specific examples of this behaviour include the noisy logistic
equation (with {\em additive} noise)
\begin{equation}
\label{E:logistic}
{\d\over\d t} \phi = 
r \phi \left( 1- {\phi\over\phi_*}\right) + \eta \; ,
\end{equation}
and the noisy Lotka-Voltera equation (used as a model for
predator-prey interactions)
\begin{eqnarray}
{\d\over\d t} N &=& N (a + b P) + \eta_N\; ,
\\
{\d\over\d t} P &=& P (c N - d)+ \eta_P \; .
\end{eqnarray}
More generally, one can easily construct noisy versions of standard
toy models as the Michaelis-Menten model for enzymatic autocatalytic
reactions, the Goodwin switch (a model of feedback control), the
Brusselator, the Fitzhugh-Naguno model of nerve potentials, or the
Field-Noyes model for oscillating reactions~\cite{Murray}.  }

{In $d=0$ there} are tremendous simplifications in the general
formalism. For time translation-invariant Gaussian noise it follows
that
\begin{eqnarray}
{\cal V}[\phi;\phi_0;d=0] &=& 
\half P^2(\phi)
+ \half \A \int {\d \omega\over (2\pi)}
\ln  
\left[ 1 + {\tilde g_2{}(\omega) 
P(\phi) P''(\phi)
\over 
\omega^2 + [P'(\phi)]^2}
\right]
- \left( \phi \to \phi_0 \right)
+ O(\A^2) \; .
\end{eqnarray}
Equivalently,
\begin{eqnarray}
{\cal V}[\phi;\phi_0;d=0] &=& 
\half P^2(\phi)
+ \half \A \int {\d \omega\over (2\pi)}
\ln  
\left[ {
\omega^2 + [P'(\phi)]^2 + 
\tilde g_2{}(\omega) 
P(\phi)  P''(\phi) 
\over 
\omega^2 + [P'(\phi)]^2}
\right]
- \left( \phi \to \phi_0 \right)
+ O(\A^2) \; .
\end{eqnarray}
Just as in the case of field theory ($d\geq 1$) this is as far as we
can go without making any further assumptions about the noise.  For
instance, {\em temporally white} noise implies $\tilde g_2(\omega)\to
1$. Integrating over frequencies and using the integral
(\ref{E:frequency}) supplemented by (\ref{E:frequency2}) and
(\ref{E:frequency3}), yields
\begin{eqnarray}
{\cal V}[\phi;\phi_0;d=0] &=& 
\half P^2(\phi) 
+ \half \A 
\left\{
\Re\sqrt{ [P'(\phi)]^2 +P(\phi)  P''(\phi) } 
- \sqrt{  [P'(\phi)]^2  }
\right\}
- (\phi \to \phi_0)
+ O(\A^2)
\; ,
\end{eqnarray}
which can be re-written as
\begin{eqnarray}
\label{E:dzero}
{\cal V}[\phi;\phi_0;d=0] &=& 
\half P^2(\phi) 
+ \half \A 
\left\{ \Re\sqrt{ \half[P^2(\phi)]'' } - |P'(\phi)|  \right\}
- (\phi \to \phi_0)
+ O(\A^2) \; .
\end{eqnarray}
Note that from a field theory point of view a SDE in $0+1$ dimensions
is {\em almost} quantum mechanics, and one-loop physics is {\em
almost} semi-classical JWKB physics. {To see what we mean by this,
consider the quantum mechanics of a system governed by the Lagrangian
\begin{equation}
L = \half m \left({\d x\over\d t}\right)^2 - V(x)\; .
\end{equation}
The techniques more usually applied to quantum {\em field theory} can
also be applied to quantum {\em mechanics} to obtain a quantum
mechanical effective potential
\begin{eqnarray}
\label{E:dzero-QM}
{\cal V}_{\mathrm QM}[x;x_0] &=& 
\half V(x) 
+ \half \hbar \;\Re\left[ \sqrt{ {V''(x)\over m} }  \right]
- (x \to x_0)
+ O(\hbar^2) \; .
\end{eqnarray}
This effective potential has the standard interpretation of being the
minimum expectation value of the Hamiltonian operator, when extremized
over stationary states satisfying $\langle \hat x \rangle = \vec
x$. In this case there is a second possible interpretation in terms of
the zero-point energy associated with the natural oscillation
frequency $\Omega = \sqrt{V''/m}$, and it is in this sense that
one-loop quantum mechanics is equivalent to semi-classical quantum
mechanics. The reaction-(diffusion)-decay system is formally very
similar to one-loop quantum mechanics with the replacement $V(x) \to
\half P(\phi)^2$; the only difference arising from the manner in which
the Jacobian is treated (the $|P'(\phi)|$ term). In quantum mechanics
two classically degenerate minima often have their degeneracy {\it
broken via semi-classical effects}: from the discussion above,
analogous phenomenon is seen to occur in stochastic mechanics.

(Note, however, that this is an analogy, not an identity. The SDE
always gives rise to Wiener functional integrals, analogous to
Euclidean quantum mechanics, instead of the Feynman functional
integrals of quantum mechanics, and the SDE never exhibits the
interference phenomena and complex amplitudes that are so
characteristic of quantum mechanics.)}

\subsection{Reaction-diffusion-decay: $d=1$}

{Spatial structures and patterns in one dimension, such as for
example the prenatal tail markings in {\it Genetta genetta} (common
genet) can be successfully modeled by one-dimensional
reaction-diffusion equations. The reader is referred to the book by
Murray for more concrete examples~\cite{Murray}.}

For $d=1$ the relevant integral, though finite, is not analytically
tractable. We are interested in evaluating
\begin{eqnarray}
\label{E:p1}
{\cal V}[\phi;\phi_0;d=1] &=& 
\half P^2(\phi) 
+ {1\over2\pi} \A \int_0^\infty {\d k} 
\left\{
\Re \sqrt{  [\nu k^2 - P'(\phi) ]^2 
+  
P(\phi) P''(\phi) 
} 
- \sqrt{ [ \nu k^2 - P'(\phi) ]^2  }
\right\}
\nonumber\\
&&
- (\phi \to \phi_0)
+ O(\A^2)\; .
\end{eqnarray}
By expanding in a power series, rescaling, integrating, and summing we
obtain the following result
\begin{eqnarray}
\label{E:series}
{\cal V}[\phi;\phi_0;d=1] &=& 
\half P^2(\phi) 
+ {1\over 2\pi} \A \; {[-P'(\phi)]^{3/2}\over\nu^{1/2}}
\sum_{n=1}^\infty
{\half\choose n} {\Gamma(1/2) \; \Gamma(2n-3/2) \over 2\;\Gamma(2n-1)}
\left({P(\phi)\;P''(\phi)\over[P'(\phi)]^2}\right)^n 
\nonumber\\
&&
- (\phi \to \phi_0)
+ O(\A^2)\; .
\end{eqnarray}
This particular expansion only makes sense for $P'(\phi)<0$, which is
the region of field configuration space in which small perturbations
die away in the absence of noise~\footnote{
Inserting $\phi(t) = \phi_0 + \delta \phi(t)$ into the noiseless
version of (\ref{E:version1}), and expanding to first order in the
time-dependent perturbation $\delta \phi(t)$ yields: $\d
[\delta\phi(t)]/\d t = P'(\phi_0) \; \delta\phi(t)+ O[(\delta
\phi)^2]$.  So for $P'(\phi_0)<0$ perturbations are damped in the
absence of noise.
}. Furthermore, the radius of convergence of the resulting sum is equal
to one, so that this expression is limited to the region $P P'' <
(P')^2$. These limitations are not fundamental, but are artifacts of
the expansion and integration procedure (the individual limits of
first expanding and then integrating term by term do not commute)
which must be taken into consideration.

\begin{figure}[htb]
\vbox{\hfil\epsfbox{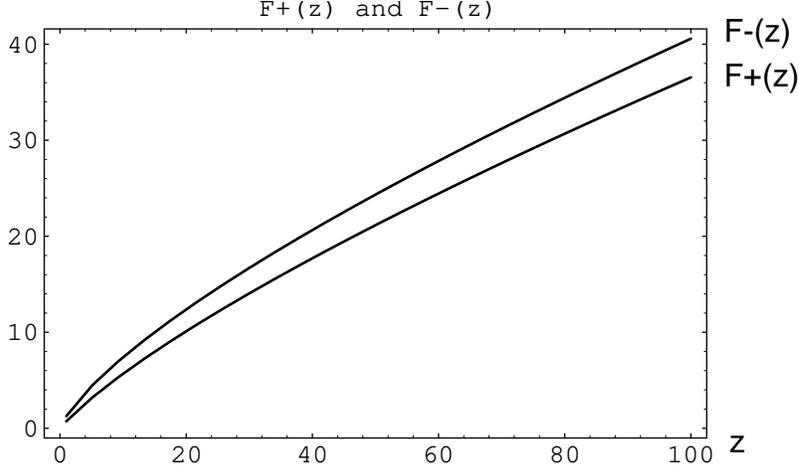}\hfil}
\bigskip
\caption{Plot of $F_\pm(z)$ for positive values of $z$, 
from zero to $100$.} 
\label{F:Fpm} 
\end{figure}

\begin{figure}[htb]
\vbox{\hfil\epsfbox{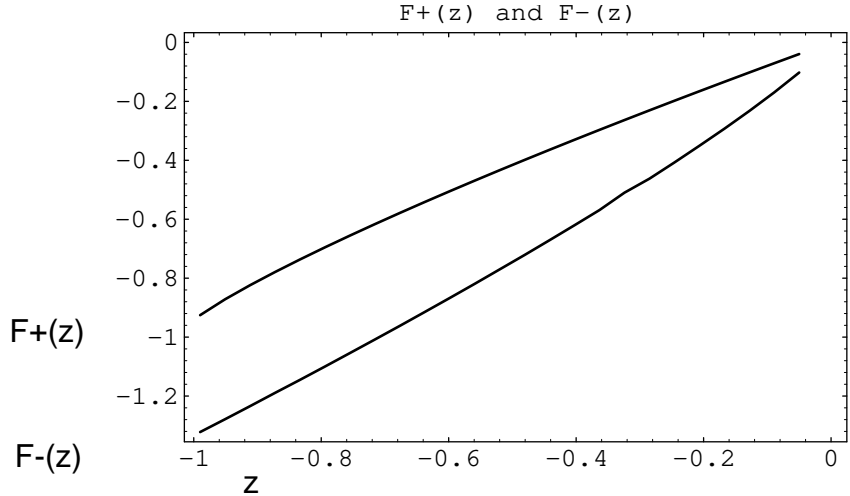}\hfil}
\bigskip
\bigskip
\caption{Plot of $F_\pm(z)$ for negative values of $z$, 
from $-1.00$ to zero.}
\label{F:Fpm2} 
\end{figure}

It is easy to see that the integral of equation (\ref{E:p1}) is real,
convergent, and well behaved for $P'(\phi)>0$ and that $P(\phi) \;
P''(\phi) > [P'(\phi)]^2$. Indeed, there is an exact (if rather
formal) representation of this integral in terms of incomplete
elliptic integrals of the first and second kinds.  The integral can
also be evaluated in terms of a ${}_3 F_2$ generalized hypergeometric
function and/or an assortment of complete Elliptic integrals~\footnote{
For further details see~\cite{G+R}, section (9.14), page 1045, and/or
section (8.1).
}. Though exact, these formulations are too cumbersome to be useful
and we can better understand the general situation by rescaling the 
original integral, using $k = \sqrt{|P'(\phi)|/\nu} \; x$, to give
\begin{eqnarray}
{\cal V}[\phi;\phi_0;d=1] &=& 
\half P^2(\phi) 
+ {1\over2\pi} \A \; {|P'(\phi)|^{3/2}\over\nu^{1/2}} \int_0^\infty {\d x} 
\left\{
\Re \sqrt{  (x^2 \pm 1)^2 
+  
{P(\phi) \; P''(\phi)\over [P'(\phi)]^2} 
} 
- |x^2 \pm 1|  
\right\}
\nonumber\\
&&
- (\phi \to \phi_0)
+ O(\A^2)
\; ,
\end{eqnarray}
where $\pm = -{\mathrm sign}[P'(\phi)]$. Thus without any detailed
calculations we know that the form of the effective potential is
\begin{eqnarray}
\label{E:done}
{\cal V}[\phi;\phi_0;d=1] &=& 
\half P^2(\phi) 
+ {1\over2\pi} \A  {|P'(\phi)|^{3/2}\over\nu^{1/2}} \; F_\pm\left[
{P(\phi) \; P''(\phi)\over [P'(\phi)]^2} 
\right]
- (\phi \to \phi_0)
+ O(\A^2)\; ,
\end{eqnarray}
with $F_\pm[z]$  the function
\begin{eqnarray}
F_\pm[z] = \int_0^\infty {\d x} 
\left\{
\Re \sqrt{  (x^2 \pm 1)^2 +  z } - |x^2 \pm 1|  
\right\} \; ,
\label{E:F+-}
\end{eqnarray}
such that $F_\pm[z=0]=0$. It is $F_+[z]$ that corresponds to the case
$P'(\phi)<0$ discussed above [see equation (\ref{E:series})]. The case
$F_-[z]$ is trickier as there is no simple Taylor series expansion
around $z=0$ (at least not in integer powers of $z$). {The overall
shape of these functions can be seen in Figures
(\ref{F:Fpm}) and (\ref{F:Fpm2}).}


{ Near a {\em stable fixed point} one has $P'(\phi)<0$, as mentioned
above. We can use either equation (\ref{E:series}) or expand equation
(\ref{E:F+-}) for $F_+(z)$ to deduce that
\begin{eqnarray}
\label{E:series-truncated}
{\cal V}_{\mathrm stable}[\phi;\phi_0;d=1] &=& 
\half P^2(\phi) 
+ {1\over 2\pi} {\A\over\nu^{1/2}}
 {\Gamma(\half)^2\over 4}
{P(\phi)\; P''(\phi)\over[-P'(\phi)]^{1/2}}
- (\phi \to \phi_0)
+ O\left(\A \; {P^2 (P'')^2 \over (P')^{5/2}} \right) 
+ O(\A^2) \; .
\end{eqnarray}
This is the universal behaviour near a stable fixed point of the PDE.

On the other hand, at a point of {\em neutral stability} for the
original noise-free PDE we have $P'(\phi)\to 0$, corresponding
to $z\to\infty$. Fortunately, the {\em large $z$} behaviour can be
analyzed analytically. Some tedious integral analysis leads to the
expressions
\begin{eqnarray}
F_\pm(z) 
&=& 
A_{+\infty} \; z^{3/4} \; \left[ 1 + O(1/z) \right]
=
{1\over6} {\Gamma(\quarter)^2\over\sqrt{\pi}} \; z^{3/4} \;
\left( 1 + O(1/z) \right)
\\
&\to& 
1.23605... \; z^{3/4} 
\left( 1 + O(1/z) \right) \qquad \hbox{as} \qquad z \to +\infty.
\end{eqnarray}
\begin{eqnarray}
F_\pm(-z) 
&=& 
- A_{-\infty} \; z^{3/4} \;
\left[ 1 + O(1/z) \right]
=
-{1\over6} {\Gamma(\quarter)^2\over\sqrt{2\pi}} \; z^{3/4} \;
\left( 1 + O(1/z) \right)
\\
&\to& 
-0.87401... \; z^{3/4} 
\left( 1 + O(1/z) \right) \qquad \hbox{as} \qquad z \to +\infty.
\end{eqnarray}
Thus near a point of {\em neutral stability} we find universal
behaviour described by
\begin{eqnarray}
\label{E:done-neutral}
{\cal V}_{\mathrm neutral}[\phi;\phi_0;d=1] = 
\half P^2(\phi) 
+ {1\over2\pi} \A  
\; A_{\pm\infty}  {[P(\phi) P''(\phi)]^{3/4}\over\nu^{1/2}} 
- (\phi \to \phi_0)
+ O\left(\A P'(\phi)^2 \sqrt[4]{P(\phi) P''(\phi)}\right) + O(\A^2)
\; .
\end{eqnarray}
Finally, near an {\em unstable fixed point} $P'(\phi)>0$. Then we must
use equation (\ref{E:F+-}) for $F_-(z)$. For $z>0$ a (not particularly
obvious) series of manipulations leads to
\begin{eqnarray}
{\d F_-(z)\over \d z}
&=&
\half \int_0^\infty {\d x \over \sqrt{(x^2-1)^2 + z}}
=
\half \Re 
\left[{1\over\sqrt{1+i\sqrt{z}}} \;
{\mathbf K} \left( \sqrt{1-i\sqrt{z} \over 1+i\sqrt{z}} \right)
\right] \; .
\end{eqnarray}
Here ${\mathbf K}(k)$ denotes a complete elliptic integral of the
first kind.  Unfortunately, although well-adapted to
numerical work, this expression is not particularly illuminating from
an analytic perspective and we do not pursue it any further.

To summarize: we see that in $d=1$ space dimensions a lot can be said
about the structure of the effective potential, without ever having to
specify the precise nature of the driving term. We trust that the
general outline of the calculation is clear, and that applying the
method to specific examples will now be straightforward.

}

\subsection{Reaction-diffusion-decay: $d=2$}

{In two spatial dimensions, reaction-diffusion equations have been
extensively employed as models for the formation of patterns on animal
coats (such as leopard spots), wing-marking patterns on
butterflies, or fingerprint development in humans,
etc.~\cite{Murray}.}

For $d=2$ the integral needed for the calculation of the effective
potential is analytically tractable. After a change of variable,
$x=\nu k^2$, and the introduction of an ultraviolet cutoff, $\Lambda$,
the desired integral becomes
\begin{eqnarray}
\int_0^{\nu \Lambda^2} \d x \sqrt{(x-P')^2 + P P''}  &=&
\half P' \sqrt{(P')^2+P P''} - \half P P'' \ln[\sqrt{(P')^2+P P''}-P']
\nonumber\\
&&
+ \half (\nu \Lambda^2-P') \sqrt{(\nu \Lambda^2-P')^2+P P''} 
\nonumber\\
&&
+ \half P P'' \ln[\sqrt{(\nu \Lambda^2-P')^2+P P''}+ \nu \Lambda^2-P']
\; .
\end{eqnarray}
This integral is tabulated in reference~\cite{G+R}, equations (2.261)
and (2.262.1). For the time being we assume that all arguments of both
the square roots and the logarithms are positive, and leave the more
technical details for later. The integral that appears in the one-loop
effective potential, equation (\ref{E:integral}), yields
\begin{eqnarray}
\int_0^{\nu \Lambda^2} \d x \left[ \sqrt{(x-P')^2 + P P''}  - 
\sqrt{(x-P')^2} \right]
&=&
\half P' \sqrt{(P')^2+P P''} - \half (P')^2 -
\half P P'' \ln[\sqrt{(P')^2+P P''}-P']
\nonumber\\
&&
+ \half (\nu \Lambda^2-P')^2 
\left[ \sqrt{1+{P P''\over(\nu \Lambda^2-P')^2}} - 1 \right]
+ \half P P'' \ln(2 \nu \Lambda^2)
\nonumber\\
&&
+ \half P P'' \ln\left[
{\sqrt{(\nu \Lambda^2-P')^2+P P''}+\nu \Lambda^2-P'\over2 \nu \Lambda^2}
\right] \; .
\end{eqnarray}
Taking the $\Lambda\to\infty$ limit, we get
\begin{eqnarray}
\int_0^{\nu \Lambda^2} \d x \left[ \sqrt{(x-P')^2 + P P''} - 
\sqrt{(x-P')^2} \right]
&=&
\half P' \; \left[\sqrt{(P')^2+P P''} - P'\right]
- \half P P'' \ln\left[\sqrt{(P')^2+P P''}-P'\right]
\label{E:lambdalimit}
\\
&&
+\half P P'' \ln (2\nu\Lambda^2) + \quarter P P''+O[1/(\nu\Lambda^2)] \; .
\nonumber
\end{eqnarray}
This explicitly verifies the presence of the logarithmic term expected
from naive power counting~\cite{HMPV-sdw-rdd}. This logarithm is the
only divergent contribution, and since it is proportional to $P(\phi)
P''(\phi)$, the (one-loop) regularization may be performed by
introducing the renormalization scale $\mu$ and making the following
split into renormalized parameters and counterterms
\begin{equation}
P_{\mathrm bare}(\phi) = P_{\mathrm renormalized}(\phi) 
+ \A \; K \; P''_{\mathrm renormalized}(\phi) \; \ln(\Lambda^2/\mu^2)
\; ,
\end{equation}
where $K$ is a calculable numerical constant whose precise value is
not important for the present discussion. It is useful to make an
additional finite renormalization in order to eliminate the $\quarter
P P''$ term from equation (\ref{E:lambdalimit}). {(It is important to
realise that these finite renormalizations do not affect the ground
state structure. They are simply equivalent to a convenient choice of
renormalization scale $\mu$.)}

After carrying out these steps, the one-loop effective
potential becomes
\begin{eqnarray}
\label{E:dtwo}
{\cal V}[\phi;\phi_0;d=2] &=& \half P^2(\phi) + {\A\over16\pi\nu}
\left[
P' \left(\sqrt{(P')^2+P P''} - P'\right) 
-  P P'' \ln\left({\sqrt{(P')^2+P P''}-P'\over\nu\mu^2}\right)
\right]
\nonumber\\
&&
- (\phi \to \phi_0)
+ O(\A^2).
\end{eqnarray}
Like the action, the one-loop effective potential cannot explicitly
depend on the renormalization scale $\mu$. In fact, the
renormalization group equation tells us that
\begin{equation}
\label{E:RGE}
\mu \frac{\d}{\d \mu}{\cal V}[\phi;\phi_0;d=2]=0\;
\Rightarrow\;
\mu{\d  P(\phi)\over \d \mu} = - {\A\over8\pi\nu} P''(\phi) +O(\A^2)
\; ,
\end{equation}
which when combined with wavefunction renormalization and the general
theorem of algebra will give the renormalization group equations for
the couplings in ${\cal V}[\phi;\phi_0;d=2]$~\cite{HMPV-tutorial}.  We
note that this equation is similar in form to the one found for the
$P(\phi)_2$ QFT in two spacetime dimensions, see~\cite{HMPV-tutorial}.

It is clear at this stage that the inclusion of a bare tadpole term
$(P_0)_{\mathrm bare}$ is essential. If there was not a
tadpole, then $P(\phi)$ would start off as $P_1 \phi+ P_2 \phi^2+ P_3
\phi^3+\cdots$, so that the lowest order term in the zero-loop
potential would be $P_1^2 \phi^2$.  On the other hand, as is
explicitly seen in equation (\ref{E:lambdalimit}), the divergent terms
are proportional to $P P'' =(P_1 \phi+ P_2 \phi^2+ P_3
\phi^3+\cdots)(2 P_2+ 6 P_3 \phi+\cdots) \to 2 P_1 P_2 \phi + \cdots$,
and there is a divergent term proportional to $\phi$ which is not
present in the tree-level effective potential.  Thus, in order to
render the theory one-loop renormalizable a tadpole term must be
included in the tree-level potential.

Notice that if $P(\phi)$ is odd [$P(-\phi) = - P(\phi)$], then the
reaction-diffusion-decay is symmetric under the following $Z_2$
symmetry transformation
\begin{equation}
\phi \to - \phi \; \; \; {\rm and} \qquad \eta \to - \eta \; .
\end{equation}
This prevents the generation of any of the even power monomial
contributions to the polynomial $P(\phi)$, including the tadpole, and
one never needs to introduce the tadpole at the tree-level.  Notice
however that this symmetry is not relevant to a realistic
reaction-diffusion-decay system, since it excludes any two-body
reactions (in fact all $2n$-body reactions).

Finally, we mention what happens when one has to deal with one of the
branch cuts that we have temporarily suppressed for simplicity of
presentation.  Although the intermediate stages of the calculation are
algebraically involved (and messy), the ultimate answer is simple:
take the real part of the expressions above.

\subsection{Reaction-diffusion-decay: $d=3$}

{An example in which prediction and observation of pattern formation
in three dimensions has created much interest is in the field of
nonlinear optical systems. Recently, three-dimensional
reaction-diffusion equations (of the Swift-Hohenberg type) have been
derived for degenerate optical parametric oscillators in which
three-dimensional Turing structures and spatial solitons have been
predicted to exist as stable structures~\cite{Staliunas}.  In addition,
morphogenesis and structural development in embryos are examples of
intrinsically three-dimensional phenomena.}

The case $d=3$ is a straightforward generalization of the $d=1$
result. We are interested in evaluating
\begin{eqnarray}
\label{E:p3}
{\cal V}[\phi;\phi_0;d=3] &=& 
\half P^2(\phi) 
+ {1\over(2\pi)^2} \A \int_0^\infty   {\d k} \; 4 \pi k^2
\left\{
\Re \sqrt{  [\nu k^2 - P'(\phi) ]^2 
+  
P(\phi) P''(\phi) 
} 
- \sqrt{ [ \nu k^2 - P'(\phi) ]^2  }
\right\}
\nonumber\\
&&
- (\phi \to \phi_0)
+ O(\A^2)\; .
\end{eqnarray}
The integral is no longer finite and a single renormalization
(without running logarithms) must be performed to absorb the infinity
in the renormalized parameters. One must be careful to keep track of
all the numerical coefficients and the final result is
\begin{eqnarray}
\label{E:dthree}
{\cal V}[\phi;\phi_0;d=3] &=& 
\half P^2(\phi)
+ {1\over (2\pi)^2} \A {[-P'(\phi)]^{5/2}\over\nu^{3/2}}
\sum_{n=1}^\infty
{\half\choose n} {\Gamma(3/2) \Gamma(2n-5/2) \over 2\Gamma(2n-1)}
\left({P(\phi)\;P''(\phi)\over[P'(\phi)]^2}\right)^n 
\nonumber\\
&&
- (\phi \to \phi_0)
+ O(\A^2) \; .
\end{eqnarray}
This expansion again only makes sense for $P'(\phi)<0$ (the region of
field space in which small perturbations die away in the absence of
noise).  Furthermore, the radius of convergence of the
resulting sum is one, so that this expression is limited to the region
$P P'' < (P')^2$. These limitations are again not fundamental, but are
merely a reflection of our choice of Taylor expansion in terms of the
field variables. {In a manner similar to what was done in $d=1$ we can
also write the above equation (\ref{E:dthree}) as
\begin{eqnarray}
\label{E:d3b}
{\cal V}[\phi;\phi_0;d=3] &=& 
\half P^2(\phi) 
+ {1\over(2\pi)^2} \A  {|P'(\phi)|^{5/2}\over\nu^{3/2}} \; F^{d=3}_\pm\left[
{P(\phi) \; P''(\phi)\over [P'(\phi)]^2} 
\right]
- (\phi \to \phi_0)
+ O(\A^2) \; ,
\end{eqnarray}
with $F^{d=3}_\pm[z]$  now being the function
\begin{eqnarray}
F^{d=3}_\pm[z] = \int_0^\infty {\d x}  \; x^2 
\left[
\Re \sqrt{  (x^2 \pm 1)^2 +  z } - |x^2 \pm 1| - \half {z\over x^2} 
\right] \; ,
\label{E:F+-d3}
\end{eqnarray}
again such that $F_\pm[z=0]=0$. It is $F_+[z]$ that corresponds to the
case $P'(\phi)<0$ discussed above [see equation (\ref{E:dthree})]. The
case $F_-[z]$ is again trickier as there is no simple Taylor series
expansion around $z=0$. Note that the last term $z/x^2$ is the
counterterm introduced to guarantee UV finiteness, and that the
integral is also IR finite. These functions can be analyzed in a
manner analogous to the discussion for $d=1$ but for the sake of
brevity we do not repeat details which are left to the industrious
reader. }

\subsection{Special cases: monomial interactions}

There are nice simplifications for monomial interactions, where
$P(\phi) = \xi \phi^n$. (Because of the generic presence of the
tadpole term this monomial behaviour should always be imposed on the
renormalized interactions, not the bare ones.) For monomial
interactions the combination $P(\phi) P''(\phi) /[P'(\phi)]^2$ reduces
to the constant $(n-1)/n$, and the one-loop effective potential (for
$d=1$, and suppressing $\phi_0$ for convenience) becomes
\begin{eqnarray}
{\cal V}[\phi;d=1] &=& 
\half \xi^2\;\phi^{2n}
+ {1\over 2\pi} \A {|-\xi\;\phi^{n-1}|^{3/2}\over\nu^{1/2}} K_\pm[1;n]
+ O(\A^2) \; ,
\end{eqnarray}
with $K_{\pm}[1;n]$ a calculable ($n$-dependent) dimensionless
constant.  If the differential equation (\ref{E:version1}) (without
noise) is assumed to be stable against small perturbations, the
coupling $\xi$ must be negative, and the exponent $n$ must be an odd
integer $n=2m+1$. In this case $P'(\phi) = (2m+1)\xi \phi^{2m} < 0
\Leftrightarrow \xi < 0$ guarantees that linear time-dependent
perturbations will decay in time. The effective potential becomes
\begin{eqnarray}
{\cal V}[\phi;d=1] &=& 
\half \xi^2\;\phi^{4m+2}
+ {1\over 2\pi} \A {|\xi|^{3/2}\;|\phi|^{3m}\over\nu^{1/2}} K_+[1;2m+1]
+ O(\A^2)\; .
\end{eqnarray}
{From} the previous equation, one can see that noise induced
corrections to the effective potential dominate for small fields in
one space dimension.

For two space dimensions the restriction to monomial interactions
implies
\begin{equation}
{\cal V}[\phi;d=2]
= \half \xi ^2 \phi^{2n} + 
{\A\over16\pi\nu} \xi^2 \phi^{2n-2}
\left\{
 n\left(\sqrt{2 n^2 -n} - n\right) 
-  n (n-1) \ln\left[{\xi\phi^{n-1}(\sqrt{2n^2-n}-n)\over\nu\mu^2}\right]
\right\}
+ O(\A^2)\; .
\end{equation}
By making a finite renormalization $\mu \to \bar \mu$, we can simplify
the previous equation to obtain
\begin{equation}
{\cal V}[\phi;\phi_0;d=2]
= \half \xi ^2 \phi^{2n} - 
{\A\over16\pi\nu} \; \xi^2 \; \phi^{2n-2}  \;
n (n-1) \ln\left({\xi\phi^{n-1}\over\nu\mu^2}
\right)
+ O(\A^2)\; .
\end{equation}
The coefficient of the logarithm is negative, implying a breakdown of
perturbation theory for large fields. More importantly, since the
coefficients of $P(\phi)$ run at one-loop, and a monomial is not a
fixed point of the renormalization group equations, if we tune the
interaction to be monomial at some fixed scale $\bar \mu$ then (in
$d=2$) the interaction will not remain monomial if the scale is
changed.

For three space dimensions the situation is similar to $d=1$.  For a
monomial interaction the one-loop effective potential is given by
\begin{eqnarray}
{\cal V}[\phi;\phi_0;d=3]
&=& 
\half \xi^2\;\phi^{2n}
+ {1\over (2\pi)^2} \A {|-\xi\;\phi^{n-1}|^{5/2}\over\nu^{3/2}} K_\pm[3;n]
+ O(\A^2)\; .
\end{eqnarray}
For a system stable in the absence of noise ($\xi<0$ and $n=2m+1$), we
have
\begin{eqnarray}
{\cal V}[\phi;\phi_0;d=3]
&=& 
\half \xi^2\;\phi^{4m+2}
+ {1\over (2\pi)^2} \A {|\xi|^{5/2}\;|\phi|^{5m}\over\nu^{3/2}} K_+[3;2m+1]
+ O(\A^2)\; .
\end{eqnarray}
In this case the noise induced effects (one-loop effects) become
important for strong fields.

\subsection{Special cases: higher dimensions}

We have seen that arbitrary polynomials in the
reaction-diffusion-decay system subject to white noise are one-loop
ultraviolet renormalizable in 0, 1, 2, and 3 space dimensions (in
fact, finite for 0 and 1 space dimensions).  We can extend the range
of dimensionalities in which these systems are one-loop renormalizable
at the cost of restricting the form of the interaction.

For instance, in four space dimensions we have already seen that there
is a divergence proportional to $P(\phi) P'(\phi) P''(\phi)$.  For
general $P(\phi)$ this cannot be renormalized, but if $P(\phi)$ is a
polynomial of degree three or less, then $P'(\phi) P''(\phi)$ is also
a polynomial of degree three or less. For this restricted class of
interactions, the divergence can be absorbed into bare potential even
in four space dimensions.

In six space dimensions there are two new divergent terms. They are
proportional to $P(\phi) [P'(\phi)]^2 P''(\phi)$ and $[P(\phi)
P''(\phi)]^2$, respectively.  If $P(\phi)$ is a polynomial of degree
two or less then $[P'(\phi)]^2 P''(\phi)$ is also a polynomial of
degree two or less and the renormalization program can be carried
out. In this case $P''(\phi)$ is either a constant or zero, so that
the second type of divergence is no further obstruction.

Finally, in eight space dimensions there is only one new divergent
term. It is proportional to $P(\phi) [P'(\phi)]^3 P''(\phi)$ and so
the theory is one-loop renormalizable only for linear interactions
({\em i.e.,} for free fields where the theory is not only
renormalizable but is actually finite.)

\subsection{More general noise}

It is clear from the above arguments that the ultraviolet
renormalizability of the reaction-diffusion-decay system depends
critically on the large momentum behaviour of the noise two-point
function.  In some problems, colored noises maybe of interest, and we
just give a necessarily brief discussion of the modifications needed
in our analysis.  Let us suppose that the noise is more general than
(space-time) white noise.  For instance, let us assume the noise is
still temporally white, but spatially power-law distributed in the
ultraviolet region with
\begin{equation}
\tilde g_2(\vec k) =\tilde g_2(\vert \vec k \vert) 
\approx
(k/k_0)^{-\theta} \; \Theta(\Lambda-k) \; ,
\end{equation}
where the positive exponent $\theta$ characterizes the strength of the
ultraviolet-singular noise.  In order to obtain the divergence
structure of the one-loop effective potential, we must make use of
equation (\ref{E:integral}). It is then easy to see that the first two
terms in the expansion for the effective potential have ultraviolet
behaviour proportional to $P(\phi) P''(\phi)\; \Lambda^{d-2-\theta}$,
and $P(\phi)\; P'(\phi) \; P''(\phi) \; \Lambda^{d-4-\theta}$,
respectively. Since the bare potential is $P^2(\phi)$, to have any
hope of absorbing the infinities into the bare action we must have
$d<4+\theta$.  The one-loop effective potential for the
reaction-diffusion-decay system is then one-loop ultraviolet
renormalizable for $d<4+\theta$ spatial dimensions.

If the noise is not temporally white, (but still Gaussian), one must
revert to equation (\ref{E:rdd-general-noise}) and perform a
case-by-case study.

\section{Noise and stability: An application}

In this section we illustrate in broad strokes how the one-loop
effective potential can be used to investigate the onset of
instabilities and pattern formation in physical, biological, and
chemical systems~\cite{Murray,Walgraef,ball} modelled by reaction-diffusion
equations (\ref{E:version1}).

By way of a concrete example which will serve as a simple template
for the one-loop equation, we consider the following model:
\begin{eqnarray}
\label{E:toy}
\frac{\partial \phi}{\partial t} - \nu \vec \nabla^2 \phi &=& P(\phi) +
\eta(\vec x, t) \; ,
\end{eqnarray}
where $P(\phi) = a\phi^2 + b\phi +c$ is the kinetic reaction
polynomial parameterized by three real constants $a, b, c$, and the
diffusion constant $\nu$ is a positive real number.  We take the noise
to be Gaussian and white.  We first briefly run through the standard
steps needed to perform a linear stability analysis of the noiseless
or zero-loop version\footnote{
The noiseless limit of general reaction-diffusion equations need
not necessarily coincide with the zero-loop limit, though these
limits are in fact identical for our model equation; see the further
comments to this effect below. For general details regarding the
distinction between no-noise and zero-loops, see reference~\cite{HMPV-spde}.
} 
of (\ref{E:toy}). This will serve as a point of reference when we
come to discuss the linear stability analysis to be performed on
the effective one-loop version of (\ref{E:toy}).
 
\subsection{Zero-noise analysis}

As is well known, the study of the onset of symmetry breaking
instabilities, be they Hopf bifurcations or Turing instabilities,
starts by classifying all the static and spatially homogeneous
solutions $\phi_0$ of the reaction-diffusion equation at hand. These
constant field configurations represent the maximally symmetric states
of the system, which could be stable or unstable with respect to time
and/or space dependent disturbances. For our toy model, these states
satisfy
\begin{equation}
\label{E:condition}
P(\phi_0) = 0\; ,
\end{equation}
which is the result of evaluating (\ref{E:toy}) for constant fields
in the absence of noise (which is completely equivalent to the
zero-loop limit of this equation). The solution is immediate, namely
we have that
\begin{equation}
\label{E:constants}
\phi_0^{\pm} = \frac{-b \pm \sqrt{b^2 - 4ac}}{2a}\; .
\end{equation}
It should be pointed out that if the field $\phi$ represents a
chemical concentration, then $\phi$ and $\phi_0$ must be non-negative.
There are many choices of the control parameters $a,b,c$ for which
this condition is met.  With knowledge of the constant states, the
next step is to expand about them so that we can study both the
temporal and spatial evolution of disturbances with respect to these
symmetric states.  Setting $\varphi^{\pm} = \phi - \phi_0^{\pm}$,
leads to
\begin{equation}
\label{E:perturbations}
\frac{\partial \varphi^{\pm}}{\partial t} - 
\nu \vec \nabla^2 \varphi^{\pm} =
\pm \varphi^{\pm}\;  \sqrt{b^2 - 4ac} + a (\varphi^{\pm})^2\; ,
\end{equation}
which is the {\em exact} zero-noise equation for the perturbations.
In arriving at this expression, we have used (\ref{E:condition}) and
$P'(\phi_0^{\pm})= {\pm}\sqrt{b^2 - 4ac}, P''(\phi_0^{\pm}) = 2a$.
To study the onset of linear instabilities, we need only focus on
the linear part of this equation. This is most conveniently carried
out in momentum or mode space, for which we introduce the Fourier
transform (in any number $d$ of spatial dimensions) of the fluctuation
$\varphi (\vec x, t)$
\begin{equation}
\label{E:Fourier}
\varphi(\vec x, t) = \int {\frac{\d^d {\vec q}}{(2\pi)^d}}\,
e^{i\vec q \cdot \vec x}\, \tilde \varphi_{\vec q}(t)\; .
\end{equation}
In terms of the mode functions $\tilde \varphi_{\vec q}(t)$, the
(exact) zero-noise equation for the perturbations takes the form
\begin{eqnarray}
\label{E:modes}
\frac{\partial \tilde \varphi_{\vec q}}{\partial t} 
&=& 
\left[-\nu q^2 \pm \sqrt{b^2 - 4ac} \right]\; \tilde \varphi_{\vec q} 
+ 
a \; \int {\frac{\d^d {\vec k}}{(2\pi)^d}}\; 
\tilde  \varphi_{\vec k} \; \tilde \varphi_{\vec q-\vec k} 
\nonumber \\ 
&=& \lambda(q^2)\; \tilde \varphi_{\vec q}
+ {\rm non-linearities} \; ,
\end{eqnarray}
which identifies the momentum-dependent eigenvalue $\lambda(q^2)$ and
the non-linear mode-mode coupling terms.

\subsubsection{Onset of Hopf bifurcations}

By definition, Hopf bifurcations are spatially homogeneous but time
dependent linear instabilities~\cite{Murray}. Spatial homogeneity
corresponds to the zero-momentum mode $(\vec q = \vec 0)$, so the
onset of this class of instability is revealed by studying the
properties of the zero-momentum eigenvalue: $\lambda(q^2=0)$. A
Hopf instability occurs when the real part ${\Re} \lambda(0) = 0$.
It is useful to consider the mode eigenvalues as functions of the
control parameters $(a,b,c)$ in the reaction kinetics polynomial.
If we define $b_{\rm Hopf}$ by $b_{\rm Hopf}^2 = 4ac$, then it is
easy to see that for $b<b_{\rm Hopf}$, the eigenvalue $\lambda(0)$
is pure imaginary and the linearized perturbations oscillate about
the constant and static field configurations $\phi_0^+$ and $\phi_0^-$
with the same frequency $\omega = |b^2 - 4ac|^{\frac{1}{2}}$.  On
the other hand, if $b>b_{\rm Hopf}$, then the linearized perturbations
about $\phi_0^-$ decay exponentially, and they grow exponentially
about $\phi_0^+$. The instability therefore sets in at $b=b_{\rm
Hopf}$, and only the $\phi_0^+$ field configuration is unstable.

\subsubsection{Onset of Turing instabilities}

The onset of spatial structure formation (Turing instabilities)
occurs whenever the eigenvalue satisfies the condition ${\Re}
[ \lambda(q^2) ] > 0$, for some non-zero mode or modes~\cite{Murray}.
The corresponding mode (or modes) set the length scale $L$ (or scales)
that characterize the spatial structures: $L \sim 1/q$.  Note that for
$b<b_{\rm Hopf}$, the real part of the eigenvalue for all modes is
negative, so no spatial structure can form, and this holds for both
the initial configurations $\phi_0^{\pm}$.  Furthermore for
$\phi_0^-$, $b>b_{\rm Hopf}$, and for any $\vec q$, the eigenvalue
$\lambda(q^2)$ is a negative real number.  Therefore, no spatial
patterns can develop as linear perturbations of this field
configuration.

On the other hand, consider the case $\phi_0^+$ and  $b>b_{\rm
Hopf}$. The eigenvalues for all modes are real, and of these, there
is a finite band of momentum scales for which the eigenvalue is
strictly positive:  namely, for $0 \leq q^2 < q^2_T \equiv
\sqrt{b^2-4ac}/\nu$.  So, one expects onset of spatial structures
to form with length scales corresponding to this momentum band.

This completes the purely linear stability analysis of the toy
reaction-diffusion equation (\ref{E:toy}) {\it in the absence of
noise}.

\subsection{One-loop analysis}

We now demonstrate, (by making use of the effective potential
calculated in previous sections), how the inclusion of noise at
one-loop impacts on the linear stability analysis illustrated above for
our simple reaction-diffusion model. The idea is to repeat the above
steps, but working now with the effective one-loop reaction-diffusion
equation. To obtain the equations of motion in the presence of noise,
we refer to the discussion in Appendix B of~\cite{HMPV-spde}. The
complete effective action for reaction-diffusion systems in the
presence of (arbitrary) Gaussian noise is given by
\begin{equation}
\label{E:effaction}
\Gamma[\phi;\phi_0] = \frac{1}{2} \int \int 
\d^d \vec x\; \d t \;  \d^d \vec y\; \d t' \;  
(D_{\rm eff}\phi -P_{\rm eff}[\phi]) g_2^{-1}(D_{\rm eff}\phi 
-P_{\rm eff}[\phi])\; ,
\end{equation}
where $D_{\rm eff}$ and $P_{\rm eff}$ are an effective differential
operator, and effective kinetic reaction function, (not necessarily a
polynomial!), respectively.  The full (to all loops) equation of
motion follows from the condition~\cite{HMPV-spde}:
\begin{eqnarray}
\label{E:eqnmotion1}
\frac{\delta \Gamma[\phi;\phi_0]}{\delta \phi} =
0 \; \; \; \; \;  {\rm or} \; \; \; \; \; 
\frac{\delta {\cal V}[\phi;\phi_0]}{\delta \phi} = 0\; , 
\end{eqnarray}
where the second expression holds for the effective potential and
yields the dynamical equation satisfied by static and homogeneous
field configurations.  As a quick check of the formalism, consider the
zero-loop effective potential: ${\cal V}[\phi;\phi_0] = \frac{1}{2}
P^2[\phi]$. From (\ref{E:eqnmotion1}) we must have $P(\phi)P'(\phi) =
0$, so that {\em either} $P(\phi) = 0$ and/or $P'(\phi) = 0.$ In fact,
we know from (\ref{E:condition}) that it must be $P(\phi)$ that
vanishes to correctly yield the solutions of (\ref{E:toy}) in the
noise-free static and homogeneous field limit. The ``spurious'' factor
$P'(\phi)$ is a consequence of the quadratic nature of the effective
action (\ref{E:effaction}).  It is easy to verify that $P'(\phi) \neq
0$ evaluated at the zeroes of $P(\phi)$.  This is important in that we
do not generate more solutions than those corresponding to the
zero-noise equation (\ref{E:toy}): zero-loops should correspond to
zero-noise, and we find that this does in fact hold for our model. In
general, spurious solutions will be absent whenever a certain
differential operator $\big( D - P'[\phi]\big)$ is non-singular
(invertible). Even in those situations where it can become singular, a
simple limiting procedure can be invoked to eliminate the spurious
solutions~\cite{HMPV-spde}.

Under these conditions, {\em i.e.,} making use of the invertibility
of $(D-P'[\phi])$, the one-loop equation of motion associated with
(\ref{E:toy}) is
\begin{eqnarray}
\label{E:oneloop}
\frac{\partial \phi}{\partial t} - \nu \vec \nabla^2 \phi 
+ O(\A  \vec \nabla^2 \phi, \A  \partial_t^2 \phi, \dots)
&=& 
(2{\cal V}[\phi;\phi_0])^{\frac{1}{2}}
= \left[ P^2(\phi) + 2 \A X(\phi) + O(\A^2) \right]^{\frac{1}{2}}
\; ,
\end{eqnarray}
where the one-loop terms in the effective potential denoted above by
$X(\phi)$ have been computed for general kinetic reaction functions
and may be written down by inspection from (\ref{E:dzero}),
(\ref{E:done}), (\ref{E:dtwo}), and (\ref{E:dthree}) above. It should
be noted that the one-loop contribution to the effective potential
vanishes whenever $P(\phi)=0$ so that without loss of generality we
can write the $O(\A)$ terms as
\begin{equation}
X(\phi) =  P(\phi) \; h(\phi)\; ,
\end{equation}
with $h(\phi)$ remaining finite (and generally non-zero) as one
approaches solutions of the zero-noise equations of motion.
Furthermore, at one-loop we know from (rather general considerations
detailed in~\cite{HMPV-sdw-rdd}) that the one-loop effective action
for any reaction-diffusion equation driven by white noise has no
wavefunction renormalization in fewer than six spatial dimensions.
This means that in dimension six or less, the differential operator
$D$ {\it per se} remains unchanged at one-loop. There can, however, be
{\em finite} renormalizations that induce finite and calculable
one-loop corrections that involve combinations of fields and
derivatives; we summarize the first few possible structures belonging
to this class above.  It is clear from (\ref{E:oneloop}) that as the
noise amplitude $\A$ is taken to zero, we recover the zero-loop
(noiseless) equation (\ref{E:toy}) of motion.

In view of the above, the one-loop equations of motion read
\begin{eqnarray}
\label{E:oneloop2}
\frac{\partial \phi}{\partial t} - \nu \vec \nabla^2 \phi 
+ O(\A  \vec \nabla^2 \phi, \A  \partial_t^2 \phi, ...)
&=& 
P(\phi)
\left[1 + \A { 2 h(\phi) \over P(\phi) } +O(\A^2) \right]^{\frac{1}{2}}
= P(\phi) + \A \; h(\phi) +O(\A^2)\; .
\end{eqnarray}
We are now ready to proceed with the stability analysis for the
one-loop reaction-diffusion equation (\ref{E:oneloop}). Just as for
the noiseless, zero-loop case, one begins by solving for all the
possible static and spatially homogeneous configurations.  These
will be the solutions of
\begin{equation}
\label{E:condition2}
P(\hat \phi_0) + \A \; h(\hat \phi_0) + O(\A^2)= 0\; ,
\end{equation}
no matter how complicated the derivative structure on the left-hand
side of the one-loop equations of motion may be.  We denote by ${\hat
\phi_0}$ the one-loop constant field configurations to distinguish
them from their tree-level counterparts. In fact, it is easy to
demonstrate that the zero-loop solutions $\phi_0$ are in general {\em
not} solutions of this one-loop equation (\ref{E:condition2}).  Thus,
we can conclude that typically $\hat \phi_0 \neq \phi_0$. Physically,
this reflects the fact that the presence of noise has altered the
symmetric states of the system.  Suppose we have chosen a particular
spatial dimension and have catalogued these new symmetric states. [To
do so in actual practice requires selecting one of (\ref{E:dzero}),
(\ref{E:done}), (\ref{E:dtwo}), and (\ref{E:dthree}), and solving the
resulting (algebraic-transcendental) equation implied by
(\ref{E:condition2}).  However, bear in mind that the point we wish to
make can be achieved without doing so explicitly.]  The next step
involves expanding the one-loop equation in (linear) perturbations
about these one-loop states. We define $\hat \varphi = \phi - \hat
\phi_0$, and write
\begin{eqnarray}
\label{E:perturbations2}
\frac{\partial \hat \varphi}{\partial t} - 
\nu \vec \nabla^2 \hat \varphi  
+ O(\A  \vec \nabla^2 \hat \varphi, 
\A  \partial_t^2 \hat \varphi, ...)
=
P(\hat \varphi + \hat \phi_0) 
+ 
\A \; h(\hat \varphi + \hat \phi_0) 
+
O(\A^2)\; .
\end{eqnarray}
We Taylor expand the right-hand side of (\ref{E:perturbations2})
making use of (\ref{E:condition2}). We then obtain the linearized
one-loop equation of motion
\begin{equation}
\frac{\partial \hat \varphi}{\partial t} - 
\nu \vec \nabla^2 \hat \varphi 
+ O(\A  \vec \nabla^2 \hat \varphi, \A  
\partial_t^2 \hat \varphi, \dots)
= 
\left[
P'(\hat \phi_0) +\A \; h'(\hat \phi_0)
\right] \, {\hat \varphi} + O(\A^2) + {\rm non-linearities}\; .
\end{equation}
After Fourier transformation the evolution of the mode functions
is governed by
\begin{equation}
\frac{\partial {\tilde {\hat \varphi}}_{\vec q}}{\partial t} 
= 
\hat \lambda(q^2)\;
{\tilde {\hat \varphi}}_{\vec q}
+ O(\A^2) + {\rm nonlinearities} 
-  O(\A  \partial_t^2 {\tilde {\hat \varphi}}_{\vec q}, \dots)\; ,
\end{equation}
where the one-loop mode-dependent eigenvalues are 
\begin{equation}
\hat \lambda(q^2) 
=
-\nu q^2  + P'(\hat \phi_0)  +\A h'(\hat \phi_0) 
- O(\A  q^2,\dots)\; .
\end{equation}
The function $h$ is the one-loop contribution to the effective
potential (divided by $P$) and has been explicitly calculated (for
arbitrary $P$) in various spatial dimensions. Since it is a function
of the reaction polynomial, it is also a function of the control
parameters $(a,b,c)$ appearing in $P$. If there are finite
renormalizations leading to new derivative structures, then $\hat
\lambda$ will also depend on their respective numerical coefficients.
For the toy model considered here, this means that the one-loop mode
eigenvalues $\hat \lambda$ are also functions of the same parameters
$(a,b,c)$ that appeared in the zero-loop eigenvalues in
(\ref{E:modes}), as well as of whatever finite renormalizations are
present at one-loop.

We summarize our main point: when we investigate the conditions for
the onset of Hopf and Turing instabilities at one-loop, ${\Re} [\hat
\lambda(0)] = 0$ and ${\Re} [\hat \lambda(q^2)] > 0$, $q^2 \neq 0$, we
will obtain new conditions on the parameters that govern the onset of
whatever instabilities are present, with respect to the $O(\cal A)$
noise-altered symmetric states $\hat \phi_0$. We see that the effect
of the noise is to shift the symmetric states of the system, as well
as to change the nature of the linear instabilities that may be
induced by perturbations around these new states.

\section{Discussion}

In this paper we have made use of the formalism developed
in~\cite{HMPV-spde} and have applied it to the class of stochastic
partial differential equations subject to diffusion, decay, and
polynomial reaction kinetics.  These equations are, for example,
extensively considered in the mathematical modeling of chemical
reactions, and biological pattern
formation~\cite{Murray,Walgraef,ball,van-Kampen,Gardiner}.

We have restricted that framework to the case of white noise and have
calculated the one-loop corrections (i.e., we have taken into account
the second-order fluctuations about the static and homogeneous
solutions of these equations) to the effective potential in various
spatial dimensions and for a general polynomial reaction kinetics term
$P(\phi)$. The effective potential, which provides information about
the possible ground states of the system (which may or may not be
stable to small perturbations, see below), is calculated by functional
integral methods and we find that it is one-loop finite for zero and
one space dimensions, and one-loop renormalizable in two and three
space dimensions. By {\em finite} we mean there are no short-distance
divergences, and by {\em renormalizable}, we mean that whatever
ultraviolet divergences are present, they can be absorbed into the
parameters appearing in the original stochastic partial differential
equation. In particular, in two space dimensions, this
renormalizability leads to a set of one-loop renormalization group
equations (\ref{E:RGE}) that govern the renormalization-scale
dependence of the parameters present in the original reaction
polynomial.

In understanding the onset of spatio-temporal pattern formation in
systems out of equilibrium, it has proven extremely useful to begin
the analysis by first solving for and then classifying all the static
and spatially homogeneous states allowed by the time-dependent partial
differential equations employed to model the system in
question~\cite{Cross-Hohenberg}.  In this way, one can
straightforwardly decide whether the system will exhibit Hopf
bifurcations and/or Turing instabilities and get a handle on the
qualitative nature of the pattern expected to emerge.

This can be followed up by an amplitude analysis of the {\em
fluctuations} about these static and homogeneous states. The unstable
modes are the ones that lead to non-trivial patterns.  For
out-of-equilibrium systems coupled to noisy environments (or with
inherent internal noise) it is important to know how the stochastic
sources can alter and shift these static and homogeneous states, since
these affect the onset of the pattern-forming (linear) instabilities.
It is seen that indeed noise will affect these patterns in a way which
is computable. In fact, the calculations in this paper show how the
effects of stochastic noise on these states of reaction-diffusion
systems can be taken into account in an elegant and computationally
direct way following the general formalism developed
in~\cite{HMPV-spde}.

\section*{Acknowledgments}

In Spain, this work was supported by the Spanish Ministry of Education
and Culture and the Spanish Ministry of Defense (DH and JPM). In the
USA, support was provided by the US Department of Energy (CMP and MV).
The research of CMP is supported in part by the Department of Energy
under contract W-7405-ENG-36.  Additionally, MV wishes to acknowledge
support from the Spanish Ministry of Education and Culture through the
Sabbatical Program, to recognize the kind hospitality provided by
LAEFF (Laboratorio de \Astrofisica\ Espacial y \Fisica\ Fundamental;
Madrid, Spain), and to thank Victoria University (Te Whare Wananga o
te Upoko o te Ika a Maui; Wellington, New Zealand) for hospitality
during intermediate stages of this work.

\appendix
\section{Jacobian functional determinant in zero space dimensions}

We are interested in evaluating the following functional determinant
in zero space dimensions
\begin{equation}
{\cal J} \equiv \det\left[ (\partial_t)^n - {\d P\over\d\phi} \right]\; .
\end{equation}
If $n>1$, from the general analysis given in~\cite{HMPV-spde}, we see
\begin{equation}
{\cal J}_n = \det(\partial_t)^n\; ,
\end{equation}
whereas for $n=1$, we  get~\cite{HMPV-spde}
\begin{equation}
{\cal J}_1 =  
\det(\partial_t) \; 
\exp \left( -\Theta(0) \; {\d P\over\d\phi} \right) 
\to 
\det(\partial_t) \;
\exp \left( -{1\over2} \; {\d P\over\d\phi} \right)\; ,
\end{equation}
where we have adopted the prescription $\Theta(0)=1/2$.

\section{Jacobian functional determinant for the 
reaction-diffusion-decay system}

For the reaction-diffusion-decay system $F[\phi(\vec x,t)] = P(\phi)$
is an ultra-local function of $\phi$, most commonly a polynomial in
$\phi$.  Thus
\begin{equation}
{\delta F[\phi(\vec x,t)]\over\delta\phi(\vec y,t') } \to
P'(\phi(\vec x)) \; \delta(\vec x-\vec y) \delta(t-t')\; .
\end{equation}
The relevant trace is given by~\cite{HMPV-spde}
\begin{eqnarray}
\Tr\left[G_1 {\delta F[\phi(\vec x)]\over\delta\phi(\vec y)}\right] 
&=&
\Theta(0) \int \d t \tr{}_{\mathrm space}
\left[P'(\phi) \; \delta(\vec x-\vec y) \right]
=
\Theta(0) \; \delta^d(\vec 0) \int \d^d \vec x \; \d t \; P'(\phi)\; .
\end{eqnarray}
In contrast to the Kardar-Parisi-Zhang equation~\cite{HMPV-kpz}, the
Jacobian determinant for the reaction-diffusion-decay system is {\em
not} a field independent constant, but (adopting the prescription
$\Theta(0)= 1/2$) one has
\begin{equation}
{\cal J}_{\mathrm RDD} = 
\exp \left[-{1\over2}\; \delta^d(\vec 0) \int
\d^d \vec x \; \d t \; P'(\phi)\right]\; .
\end{equation}
Thus for the reaction-diffusion-decay field theory the functional
determinant at worst leads to extremely simple Faddeev-Popov ghosts.
There are general arguments (see for example
Zinn-Justin~\cite{Zinn-Justin} pp. 373, 307, or related comments in
Itzykson-Zuber~\cite{Itzykson-Zuber} p. 448) to the effect that terms
proportional to $\delta^d(\vec 0)$ can always be safely discarded in
dimensional regularization.  We do not want to step into the middle of
this contentious issue and merely note that we have found it more
convenient to {\em not} adopt the formal result $\delta^d(\vec 0) =
0$, and instead to explicitly carry the Jacobian along in the
calculation. Keeping the Jacobian explicit is essential to showing
one-loop finiteness in $d=1$ space dimension, a result that would
otherwise be disguised by unnecessarily discarding the precise
counterterm needed to ensure the one-loop finiteness of the theory.

This situation is in marked contrast to that for the KPZ
system~\cite{HMPV-kpz}, wherein the Jacobian determinant is a field
independent constant irrespective of how one wishes to handle the
formal result $\delta^d(\vec 0)=0$.

\section{Feynman rules for the Reaction-diffusion-decay system}

We have derived the general form of the Feynman rules in the direct
formalism, applicable to arbitrary SPDEs, in~\cite{HMPV-spde}.  There
are a number of technical simplifications for the
reaction-diffusion-decay system which make it worthwhile to present
this particular case. From the reaction-diffusion-decay stochastic
differential equation (\ref{E:version1})
\begin{equation}
\left[{\partial\over\partial t} - \nu (\vec \nabla^2-m^2)\right] \phi = 
P(\phi) + \eta\; ,
\end{equation}
we deduce the characteristic functional (partition function)~\cite{HMPV-spde}
\begin{eqnarray}
Z[J] &=& 
\int ({\cal D} \phi)\; \sqrt{{\cal J}\;{\cal J}^\dagger}
\exp\left( -{1\over2} \int \int
\left[
\partial_t \phi - \nu(\vec \nabla^2-m^2) \phi - P(\phi)
\right] 
 G_\eta^{-1}
\left[
\partial_t \phi - \nu(\vec \nabla^2-m^2) \phi - P(\phi)
\right]
\right) \;
\nonumber\\
&&\qquad
\times \exp\left( 
\int \left[J\phi\right] 
\right)\; ,
\end{eqnarray}
from which we can immediately deduce the associated Feynman rules.
Note we have opted to collect whatever terms linear in $\phi$ there
may be in the reaction polynomial $P(\phi)$ and place them on the left
hand side of the equation. Compare to the equations (\ref{E:version1})
and (\ref{E:version2}) in the introduction: the decay rate is simply
$\gamma = \nu m^2$.  The functional determinant can either be
calculated from the preceding section, or it can be included as a
Faddeev-Popov ghost term, with the same effect.  There is only one
field propagator and two vertices. The propagator is identical to
the KPZ case, but the vertices have as high an order as determined by
the highest power in $P$ (see below).  The following holds for
translationally invariant noise.

These are the basic structures (in the interacting classical action)
from which one can derive the appropriate vertices by taking the
corresponding functional derivatives:
\begin{eqnarray}
&&\hbox{Propagator:}
\nonumber\\
&&
\qquad
G_{\mathrm field}(\vec k,\omega)= 
{ G_\eta(\vec k,\omega)\over \omega^2 + \nu^2 (\vec k^2+m^2)^2}\; ,
\\
&&
\hbox{$P(\phi)$-$\phi$ vertex:} 
\nonumber\\
&&
\qquad
{ [\tilde P(\phi)](\vec k_1,\omega_1) \; \tilde\phi(\vec k_2,\omega_2)
\over (2\pi)^{d+1}} \; \;
{[-i\omega_1+\nu (\vec k_1^2 + m^2)]\over G_\eta(\vec k_1,\omega_1) } \; \;
\delta(\vec k_1 + \vec k_2) \;
\delta(\omega_1 + \omega_2) \; ,
\\
&&
\hbox{$P(\phi)$-$P(\phi)$ vertex:}
\nonumber\\
&&  
\qquad
{1\over2} \; 
{[\tilde P(\phi)](\vec k_1,\omega_1) \; [\tilde P(\phi)](\vec k_2,\omega_2) 
\over(2\pi)^{d+1} G_\eta(\vec k_1,\omega_1)} \; \;
\delta(\vec k_1 + \vec k_2) \; 
\delta(\omega_1 + \omega_2)\; .
\end{eqnarray}
For a polynomial $P(\phi)$ of order $n$, there will be up to $2n$
propagators meeting at a vertex. Calculations with these Feynman rules
involve fewer propagators and vertices than those of the
Martin-Siggia-Rose formalism~\cite{MSR,De-Dominicis-Peliti}. The
basic trade-off is this: one reduces the number of propagators and
vertices at the cost of making the vertices more complicated. As long
as one is interested in one-loop physics this cost is not
prohibitive~\cite{HMPV-spde}, and the pay-off in terms
of ease of calculation for the effective action and effective
potential is considerable.

\section{An integral}

We want to demonstrate that
\begin{equation}
\int_{-\infty}^{+\infty} \d \omega 
\ln \left( {\omega^2 -Z^2 \pm i \epsilon \over \omega^2 + Y^2} \right) = 
2\pi \left(  \pm i Z - Y \right)\; .
\end{equation}
To see this note that
\begin{eqnarray}
\int_{-\infty}^{+\infty} \d \omega 
\ln \left( {\omega^2 -Z^2 \pm i \epsilon \over \omega^2 + Y^2} \right) 
&=& 
\int_{-\infty}^{+\infty} \d \omega 
\ln \left( {|\omega^2 -Z^2| \over \omega^2 + Y^2} \right) 
\pm i \pi \int_{-Z}^{+Z} \d \omega 
=
\int_{-\infty}^{+\infty} \d \omega 
\ln \left( {|\omega^2 -Z^2| \over \omega^2 + Y^2} \right) 
\pm 2\pi i Z\; .
\end{eqnarray}
The remaining integral can be found in~\cite{G+R} equation (2.736.1)
and (2.733.1), and we can finally write
\begin{eqnarray}
\int_{-\infty}^{+\infty} \d \omega 
\ln \left( {|\omega^2 -Z^2| \over \omega^2 + Y^2} \right) 
&=&
\left. \left\{ 
\omega \ln\left( {|\omega^2 -Z^2| \over \omega^2 + Y^2} \right) 
+ Z \ln\left|{\omega+Z\over \omega-Z}\right|
- 2 Y \tan^{-1}\left(\frac{\omega}{Y}\right)
\right\} \right|_{-\infty}^{+\infty}
= - 2 \pi Y\; .
\end{eqnarray}
%

\section{The Ito calculus versus the Stratonovich calculus}

In evaluating the Jacobian functional determinant one encounters a
factor of $\Theta(0)$, which is ill-defined and must be specified by
some prescription.  The prescription which is most useful in this
context, and which we have adopted in the bulk of this paper, is the
symmetric prescription wherein $\Theta(0)$ equals ${1\over2}$. This
may be justified by a limiting procedure as described, for example, in
the text by Zinn-Justin~\cite{Zinn-Justin} (Chapter 4, pp. 69-70.)
The symmetric prescription is equivalent to adopting the {\em
Stratonovich calculus} for stochastic equations. Choosing
$\Theta(0)=0$ is equivalent to the {\em Ito calculus}. The Ito
calculus simplifies the Jacobian determinant (to unity) at the cost of
destroying equivariance under field redefinitions (the Ito calculus
explicitly breaks coordinate invariance in field space). See, for
instance, Eyink~\cite{Eyink1}, or Zinn-Justin~\cite{Zinn-Justin}.  In
this Appendix we sketch the modifications required to implement the
Ito calculus. These changes are straightforward if at times tricky
(the loss of reparameterization invariance under field redefinitions
implies that all arguments involving a change of variables must be
carefully re-assessed).

If we adopt the Ito calculus, then for any SPDE the (unrenormalized)
expression for the one-loop effective potential simplifies to
\begin{eqnarray}
\label{E:general-Ito}
{\cal V}_{\mathrm Ito}[\phi;\phi_0] &=& 
\half F^2[\phi]
+ \half \A \int {\d^d \vec k \; \d \omega\over (2\pi)^{d+1}}
\ln  
\left[ 
\left( D^\dagger(\vec k,\omega)  - {\delta F\over \delta\phi}^\dagger \right)
\left( D(\vec k,\omega) - {\delta F\over \delta\phi} \right)
+ \tilde g_2{}(\vec k,\omega)
F[\phi] {\delta^2 F\over\delta\phi\;\delta\phi}
\right]
\nonumber\\
&&
- \left( \phi \to \phi_0 \right)
+ O(\A^2)\; .
\end{eqnarray}
[Compare with equation (\ref{E:general}).]  When specialized to the
reaction-diffusion-decay system this further simplifies to
\begin{eqnarray}
\label{E:rdd-Ito}
{\cal V}_{\mathrm Ito}[\phi;\phi_0] &=& 
\half \left[ P^2(\phi) - P^2(\phi_0) \right]
+ \half \A \int {\d^d \vec k \; \d \omega\over (2\pi)^{d+1}}
\ln  
\left[ {
\omega^2 + [\nu \vec k^2 - P'(\phi) ]^2 + 
\tilde g_2{}(\vec k,\omega) P(\phi) P''(\phi) 
\over 
\omega^2 + [\nu \vec k^2 - P'(\phi_0) ]^2 + 
\tilde g_2{}(\vec k,\omega) P(\phi_0) P''(\phi_0) 
} \right]
+ O(\A^2)\; .
\end{eqnarray}
[Compare with equation (\ref{E:p1rdd}).]  Further restricted to
temporally white noise we obtain
\begin{eqnarray}
\label{E:integral-Ito}
{\cal V}_{\mathrm Ito}[\phi;\phi_0] &=& 
\half P^2(\phi) 
+ \half \A \int {\d^d \vec k  \over (2\pi)^{d}} 
\left\{
\Re\left[\sqrt{  [\nu \vec k^2 - P'(\phi) ]^2 
+ \tilde g_2{}(\vec k) 
P(\phi) P''(\phi) 
} \right] 
\right\}
- (\phi \to \phi_0)
+ O(\A^2)\; .
\end{eqnarray}
[Compare with equation (\ref{E:integral}).]

On the one hand, this looks like a tremendous simplification of
equation (\ref{E:integral}).  On the other hand, the ultraviolet
renormalizability properties are now considerably worse. For instance,
by inspection of the above it is easy to see that one can no longer
rely on a cutoff in the noise spectrum, $g_2(\vec k)$, to keep the
effective potential finite. Instead a cutoff in the momentum integral
must be introduced by hand, complicating the process tremendously.
Even if $g_2(\vec k)\to 0$ for large momenta, for $d>0$ there exists
an UV divergence proportional to $\Lambda^d
[P'(\phi)-P'(\phi_0)]$. For generic $P(\phi)$ this cannot be absorbed
into a counterterm in the zero-loop effective action. That is,
adopting the Ito calculus for the reaction-diffusion-decay system
results in a theory that is one-loop non-renormalizable for $d>0$, and
so must be viewed as an ``effective field theory''.

Thus even for $d=1$, where the Stratonovich calculus leads to a
one-loop finite result, the Ito calculus leads to complicated
expressions which obscure the underlying physics.  It is for this
reason, (the complications of dealing with non-renormalizable
effective field theories), and the fact that the Ito calculus is not
invariant under field redefinitions, that we have not further explored
the Ito calculus in this paper.

The one case where the Ito calculus gives a  simpler answer
than the Stratonovich calculus is for $d=0$, corresponding to
stochastic mechanics rather than stochastic field theory. In that case
\begin{eqnarray}
\label{E:dzero-Ito}
{\cal V}_{\mathrm Ito}[\phi;\phi_0;d=0] &=& 
\half P^2(\phi) 
+ \half \A 
\left( \Re\sqrt{ \half[P^2(\phi)]'' } \right)
- (\phi \to \phi_0)
+ O(\A^2)\; .
\end{eqnarray}
Of course, one should not be alarmed that the Ito calculus and the
Stratonovich calculus give different intermediate results---they are
different theories. Because the Ito calculus is not invariant under
field redefinitions it is generally possible to find {\em some} choice
of field variables (a reparametrization) that makes the two systems
agree with each other, but that special set of field variables may not
be the ones one naively started out with.

%

\end{document}